\documentclass[preprint,prd,letterpaper,tightenlines,nofootinbib,showpacs]{revtex4}
\usepackage{amsmath,amssymb,graphicx,mathrsfs}
\usepackage{dcolumn}
\usepackage{bm}
\newcommand{\ds}{\displaystyle}
\newcommand{\ssz}{\scriptsize}
\newcommand{\up}{\underline{p}}
\newcommand{\ux}{\underline{x}}

\newcommand{\nn}{\nonumber\\}

\newcommand{\la}{\langle}
\newcommand{\ra}{\rangle}
\newcommand{\D}{\mathscr{D}}
\newcommand{\Hb}{\beta P^0}
\newcommand{\Ht}{P^0\tau}
\newcommand{\HL}{P_L^0}

\newcommand{\ben}{\begin{displaymath}}
\newcommand{\een}{\end{displaymath}}
\newcommand{\be}{\begin{equation}}
\newcommand{\ee}{\end{equation}}
\newcommand{\bea}{\begin{eqnarray}}
\newcommand{\eea}{\end{eqnarray}}

\newcommand{\bfg}{\mbox{\boldmath $\gamma$}}

\newcommand{\bphi}{\bar{\phi}}

\newcommand{\bc}{\begin{center}}
\newcommand{\ec}{\end{center}}

\newcommand{\eqn}[1]{\label{#1}}
\newcommand{\eq}[1]{Eq.~(\ref{#1})}
\newcommand{\eqs}[1]{Eqs.~(\ref{#1})}

\begin{document}
\title{Equivalence of light-front and conventional thermal field theory}
\author{A. N. Kvinikhidze}\altaffiliation[On leave from ]{The Mathematical 
Institute of Georgian Academy of Sciences, Tbilisi, Georgia}\email{sasha.kvinikhidze@flinders.edu.au}
\author{B. Blankleider}\email{boris.blankleider@flinders.edu.au}
\affiliation{Department of Physics, Flinders University, Bedford Park, SA 5042, Australia}
\date{\today}

\begin{abstract}
It is shown that light front  thermal field theory is equivalent
to conventional thermal field theory. The proof is based on the use of spectral
representations, and applies to all Lagrangians for which
such equivalence has been proven at zero temperature. It is also
pointed out that conventional spectral functions can be used to
express light-front finite temperature free propagators. As an application of
our approach, we derive the light-front finite temperature spin $1/2$ fermion propagator
in full Dirac space. 
\end{abstract}
\pacs{ }
\maketitle

\section{Introduction.}
The light-front (LF) formulation of quantum field theory, where quantization
arises from
equal LF "time" commutation/anticommutation relations, offers
some advantages over the conventional formulation where equal usual times are 
used for quantization. In this respect,  perhaps the three most often quoted 
advantages of the LF approach are: kinematical boosts, a much less complicated 
vacuum, and much simpler eigenstates. The last two properties are related and 
may be especially beneficial for thermal field theory where the main object of 
interest, the ensemble average, is just a sum over eigenstates.

With such considerations in mind, the problem of formulating thermal field 
theory on the light front has been addressed in a number of recent papers 
\cite{brodsky,das,weldon1,weldon2}. In Ref.\ \cite{das},  Alves,  Das, and 
Perez proposed the LF version of the imaginary time scalar particle propagator, 
and used this to calculate the self-energy loop diagram. The result of this 
calculation looked so different from the conventional one, that only after 
Weldon \cite{weldon2} made use of a clever transformation of the integration 
momentum variable, did it become clear that this difference is illusive.

In the present note we show that the {\em whole} LF approach using the
imaginary time scalar particle propagator proposed in Ref.\
\cite{das}, is equivalent to the conventional one. Our method is based on 
spectral representations of Green functions, and can be applied
to the case of any Lagrangian for which such equivalence has been shown 
at zero temperature
\cite{equival,rohrlich,chang1,chang2,yan,tomboulis,eyck,mccartor}.
In this respect, we note that the spectral function of the Lehmann representation  is
already well recognized as being very useful for relating
different types of Green functions (imaginary
time, real time, advanced, retarded, etc.) in the conventional approach. 
Here we show how to use the conventional spectral function to 
derive a LF free particle propagator of arbitrary spin in either imaginary or real time 
formalism. This allows us to derive fermion propagators in the full 
Dirac index space.

\section{General proof of equivalence of light front and conventional 
thermal field theory}
The problem of calculating the ensemble
average $\mbox{Tr}\, e^{-\beta \HL }{\cal O}_L$ in LF quantized
field theory\footnote{Note that the correct definition of ensemble average has energy operator $P^0_L$ in the exponential $e^{-\beta \HL }$ \cite{Alex} and not the Hamiltonian $P^-_L$ (see the discussion in Ref.\ \cite{das}).}
 can be reduced to the calculation of the fully dressed LF
imaginary time propagator $\Delta^L$ given in coordinate and momentum 
space by  (we use units where $\hbar=k_B=1$)\footnote
{To distinguish between LF and usual equal time quantization we
work in the operator rather than path integral formalism.}
\begin{subequations}\label{danger}
\begin{eqnarray}
\Delta^L(\tau,\ux)&=&(\mbox{Tr}\,e^{-\beta \HL })^{-1}
\mbox{Tr}\left\{e^{-\beta \HL }T_\tau [e^{\HL \tau}\phi(0,\ux)
e^{-\HL \tau}\bar\phi(0)]\right\},\eqn{dangera}\\
\Delta^L(i\omega_n,\up)&=& \frac{1}{\sqrt{2}}
\int_0^\beta d\tau d\ux\, e^{i(\omega_n\tau-\up\cdot\ux)}\,
\Delta^L(\tau,\ux), \eqn{dangerb}
\end{eqnarray}
\end{subequations}
where $\omega_n=2n\pi T$ for bosons, 
$\omega_n=(2n+1)\pi T$ for fermions,  $\ux=(x^-,x^\perp)$, $\up=(p^+,p^\perp)$, 
 $x^\pm=\frac{1}{\sqrt{2}}(x^0\pm x^3)$, $p^\pm=\frac{1}{\sqrt{2}}(p^0\pm p^3)$,
$\up\cdot\ux=-p^+x^-+p^\perp x^\perp$,  $\phi(x)$ is a noninteracting
particle field operator, $\phi(0,\ux)=\phi(x)|_{x^+=0}$, and where the energy operator $\HL$ and the operator of some physical quantity 
${\cal O}_L$ are defined in the LF quantized theory, i.e., they depend on interacting field operators whose commutation relations are given on the $x^+=0$ hyperplane (see Appendix A for more details).
The imaginary
time ordering product in \eq{dangera} is defined as
\be
T_\tau [e^{\HL \tau}\phi(0,\ux)e^{-\HL \tau}\bar\phi(0)]=
\theta(\tau)e^{\HL \tau}\phi(0,\ux)e^{-\HL \tau}
\bar\phi(0)\pm\theta(-\tau)\bar\phi(0)e^{\HL \tau}\phi(0,\ux)
e^{-\HL \tau}        \eqn{tauord}
\ee 
where the upper (lower) sign is for the case of bosons (fermions).
Note that although $\tau$ is associated with the usual time, in the sense that it is 
combined with $\HL$ (rather than with $P^-_L$) in the exponent of $e^{\HL \tau}$, 
\eq{tauord} nevertheless
represents the ordering of the interacting fields with respect to the 
imaginary LF ``time''  [see \eq{x'}, the discussion below \eq{x'}, and the text after Eq.\ (B5)], in 
contrast to the usual time ordering in \eq{timeord}.
It is also important to note that \eqs{danger} define the LF
imaginary time formalism exactly, and in the case of scalar particles (only), 
they define perturbation theory with the free propagator suggested in 
Ref.\ \cite{das}. As the perturbation theory for the dressed propagators of \eqs{danger} may not be immediately apparent (the exponents involve the energy operator $P^0_L$ rather than the Hamiltonian $P^-_L$), we outline its derivation in the Appendix B.

In this paper we shall prove that the analytic continuation of the fully dressed LF 
imaginary time propagator $\Delta^L(i\omega_n,\up)$ to real
energies, $i\omega_n\rightarrow p_0$, is 
identical to the analytic continuation of the fully dressed conventional 
imaginary time propagator $\Delta(i\omega_n,{\bf p})$, i.e., that $\Delta^L(p_0,\up)=
\Delta(p_0,{\bf p})$,
where
\begin{subequations}
\begin{eqnarray}
\Delta(\tau,{\bf x})&=&(\mbox{Tr}\,e^{-\Hb})^{-1}
\mbox{Tr}\left\{e^{-\Hb}T_\tau [e^{\Ht}\phi(0,{\bf x})e^{-\Ht}
\bar\phi(0)]\right\}, \eqn{timeord}   \\[1mm]
\Delta(i\omega_n,{\bf p})&=&\int_0^\beta d\tau d{\bf x}\,
e^{i(\omega_n\tau-{\bf p}\cdot{\bf x})}
\Delta(\tau,{\bf x}),
\end{eqnarray}
\end{subequations}
$\phi(0,{\bf x})=\phi(x)|_{x^0=0}$,\footnote{Hopefully no confusion
will arise from our not entirely consistent notation for $\phi(0,{\bf x})$ and $\phi(0,\ux)$; in particular $\phi(0,{\bf x}) \ne \phi(0,\ux)|_{\ux={\bf x}}$.}
and $P^0$ is the conventional energy operator.

We begin our proof by relating $\Delta^L(i\omega_n,\up)$ to the real time LF Green function
\begin{subequations}\label{realtim}
\begin{eqnarray}
\D^L(t,\ux)&=&(\mbox{Tr}\,e^{-\beta \HL })^{-1}
\mbox{Tr}\left\{e^{-\beta \HL }T_t [e^{i\HL t}\phi(0,\ux)e^{-i\HL t}
\bar\phi(0)]\right\},  \eqn{realtima} \\[1mm]
D^L(p^0,\up)&=&\frac{1}{\sqrt{2}}\int dt\,d\ux \,
e^{i(p^0t-\up\cdot\ux)}\D^L(t,\ux),\eqn{realtimb}
\end{eqnarray}
\end{subequations}
where $T_t$ is defined analogously to \eq{tauord} as
\be
T_t [e^{i \HL t}\phi(0,\ux)e^{-i \HL t}\bar\phi(0)]=
\theta(t)e^{i \HL t}\phi(0,\ux)e^{-i\HL t}
\bar\phi(0)\pm\theta(-t)\bar\phi(0)e^{i\HL t}\phi(0,\ux)
e^{-i\HL t}  .      \eqn{tord}
\ee 
This is done by utilizing 
the Lehmann representation which can be derived for LF Green
functions in a way similar to that for conventional Green
functions (see Ref.\ \cite{lebellac} for the conventional case):
\begin{subequations} \label{LehmanL} 
\begin{eqnarray}
D^L(p^0,\up)&=&i\int \frac{dp'_0}{2\pi}
\frac{\rho^L(p_0',\up)}{p_0-p_0'+i\eta}\pm
f(p_0)\rho^L(p_0,\up),\label{LehmanLa}  \\[1mm]
\Delta^L(i\omega_n,\up)&=&-\int\frac{dp'_0}{2\pi}
\frac{\rho^L(p_0',\up)}{i\omega_n-p_0'},\label{LehmanLb}
\end{eqnarray}
\end{subequations}
where $f(p_0)=(e^{\beta p_0} \mp 1)^{-1}$ is the distribution function for 
bosons (upper sign) or fermions (lower sign), and the LF spectral function (defined in Appendix C)
is, correspondingly,\footnote{We do not show explicit spin indices. For 
particles with spin, one should consider the field $\phi$ as a column vector,  
the field $\bphi$ as a row vector, and quantities like $\rho^L$, $D^L$,
$\Delta^L$, etc.,  as square matrices, in spin index space.}
\be
\rho^L(p_0,\up)=
\frac{(2\pi)^4}{\mbox{Tr}\, e^{-\beta \HL }}
(1\mp e^{-\beta p_0})\sum_{nm} e^{-\beta E_n} 
\delta^4(p-P_m+P_n)\la Ln|\phi(0)|Lm\ra\la Lm|\bar\phi(0)|Ln\ra.   \eqn{specfunc}
\ee
The difference between $\rho^L(p)$ and the conventional spectral function
$\rho(p)$  [see \eqs{Lehman}] is only in the eigenstates $|Lm\ra $ of the 
LF 4-momentum,  $P^\mu_L|Lm\ra=P^\mu_m|Lm\ra $\footnote{Here $P^\mu_L$ is
the operator of the 4-momentum in the LF approach whereas $P^\mu_m$ is its
eigenvalue corresponding to the state $|Lm\ra $}, which are different from 
the conventional ones (denoted by
$|m\ra $ in Ref.\  \cite{lebellac}). However,
in the free case there is no difference 
between these eigenstates and thus the free LF and conventional
spectral functions are identical. The unusual scalar product 
$p^0t-\up\cdot\ux$ in the exponent of \eq{realtimb} can be written in the
invariant form
$p^0t-\up\cdot\ux=p\cdot x'$ where $x'$ is defined by
\be
 x'_0=\frac{x^-}{\sqrt{2}}+t,\hspace{7mm}(x')^3=-\frac{x^-}{\sqrt{2}},\hspace{7mm}
(x')^\perp=x^\perp,\hspace{7mm}dt\,d\ux=-\sqrt{2}\,d^4x'  \eqn{x'}.
\ee
Writing
\be
e^{i\HL t}\phi(0,\ux)e^{-i\HL t}=\Phi_L(x'),
\ee
one finds that $\Phi_L$ is a LF Heisenberg field operator; i.e., for any four-vector $a$,
\be
\Phi_L(x+a)  = e^{iP_L\cdot a} \Phi_L(x) e^{-iP_L\cdot a} ,
\ee
with initial condition
\be
\Phi_L(0,\ux) = \phi(0,\ux).
\ee
Given that $\sqrt{2}(x')^+=t$, $T_t$ ordering in
\eq{realtima} implies LF time ordering  $T_+$, so that
\eqs{realtim} can be written in the form
\begin{subequations}
\eqn{DL}
\begin{eqnarray}
D^L(x)&=&(\mbox{Tr}\,e^{-\beta \HL })^{-1}
\mbox{Tr}\left\{e^{-\beta \HL }T_+ [\Phi_L(x)
\bar\Phi_L(0)]\right\}, \eqn{DLa} \\[1mm]
D^L(p)&=&\int d^4x \,e^{ip\cdot x}D^L(x), \eqn{DLb}
\end{eqnarray}
\end{subequations}
with the understanding that $D^L(p_0,\up)=D^L(p)$ and 
$\D^L(t,\ux)=D^L(x')$.

The well known relations, corresponding to the LF \eqs{LehmanL},  connecting conventional real and imaginary
time Green functions via the conventional spectral function $\rho(p)$ are 
\cite{lebellac,gorkov,walecka}:
\begin{subequations}
\begin{eqnarray}
D(p^0,{\bf p})&=&i\int\frac{dp'_0}{2\pi}
\frac{\rho(p_0',{\bf p})}{p_0-p_0'+i\eta}\pm f(p_0)\rho(p_0,{\bf p}),\eqn{Lehmana}\\[1mm]
\Delta(i\omega_n,{\bf p})&=&-\int\frac{dp'_0}{2\pi}
\frac{\rho(p_0',{\bf p})}{i\omega_n-p_0'},\eqn{Lehmanb}
\end{eqnarray}
 \eqn{Lehman}
\end{subequations}
where $D(p^0,{\bf p})$ is the conventional real time Green function,
defined as
\begin{subequations}
\begin{eqnarray}
D(t,{\bf x})&=&(\mbox{Tr}\,e^{-\Hb})^{-1}
\mbox{Tr}\left\{e^{-\Hb}T_t [e^{iP^0 t}\phi(0,{\bf x})e^{-iP^0 t}
\bar\phi(0)]\right\},\eqn{p1}\\[1mm]
D(p^0,{\bf p})&=&\int dt \,d{\bf x}\,
e^{i(p^0t-{\bf p}\cdot{\bf x})}D(t,{\bf x}).
\end{eqnarray}
\end{subequations}
For a perturbative treatment the propagator of \eq{p1} should be written
in the interaction representation \cite{smilga}, with
\be
\Phi(x)\equiv e^{iP^0 t}\phi(0,{\bf x})e^{-iP^0 t}=U(0,t)\phi(x)U(t,0),
\ee
\be
D(x)=(\mbox{Tr}\, e^{-\Hb})^{-1}
\mbox{Tr}\left\{e^{-\Hb_f}S^{-1}T [\phi(x)\bar\phi(0)S]\right\}
\eqn{ppert}
\ee
where $P^0_f$ is the free part of $P^0$, $T$ is the usual time ordering operator, $S=U(\infty,-\infty)$, and
\be
U(t_2,t_1)=T\exp\left\{-i\int_{t_1}^{t_2}dt \int_{-\infty}^\infty  d^3x \,\mathscr{P}^0_I(x)\right\},
\ee
$\mathscr{P}^0_I(x)$ being the interaction part of the Hamitonian density in the interaction picture.
We note the unusual appearance (for zero temperature perturbation theory) of the
 inverse S-matrix, $S^{-1}$, the source of doubled
degrees of freedom \cite{smilga}. An analogous LF interaction representation
can be written for $D^L$: 
\be
\Phi_L(x)\equiv e^{iP^-_Lx^+}\phi(0,\ux)e^{-iP^-_Lx^+}=
U_L(0,x^+)\phi(x)U_L(x^+,0),
\ee
\be
D^L(x)=(\mbox{Tr}\, e^{-\beta \HL })^{-1}
\mbox{Tr}\left\{e^{-\Hb_{Lf}}S_L^{-1}T_+ [\phi(x)\bar\phi(0)S_L]\right\},
\eqn{LFpert}
\ee
where $P^0_{Lf}$ is the free part of $P^0_L$, $S_L=U_L(\infty,-\infty)$, $P^-_L$ is the LF Hamiltonian, i.e.,
the minus component of the four-momentum operator, and
\be
U_L(\alpha_2,\alpha_1)=T_+\exp\left\{-i\int_{\alpha_1}^{\alpha_2} dx^+ \int_{-\infty}^\infty
d\ux\, \mathscr{P}^-_{LI}(x)\right\},
\ee
$\mathscr{P}^-_{LI}(x)$ being the interaction part of the LF Hamiltonian density in the
interaction representation.

\subsection{Scalar particles}

To compare the dressed real time propagators in the conventional and LF formalisms,  given in
\eq{ppert} and \eq{LFpert} respectively, we first restrict the discussion to the case of scalar particles. For scalar particles $\mathscr{P}^-_{LI}$ and $\mathscr{P}^0_I$ are the same functions of the free
field operators $\phi$ \cite{chang1}, which means that the perturbation theories for 
\eq{ppert} and \eq{LFpert} have the same vertices. The inverse S-matrix
in both \eq{ppert} and \eq{LFpert} leads to free propagators with doubled
degrees of freedom, i.e. $2\times 2$  matrices $\hat D^f$ and 
$\hat D^{Lf}$ whose $(1,1)$ element is defined by \eq{ppert} and \eq{LFpert}
in the no interaction limit:
\begin{subequations}\label{freeprop}
\begin{eqnarray}
\hat D_{11}^f(x)&=&D^f(x)=(\mbox{Tr}\, e^{-\Hb_f})^{-1}
\mbox{Tr}\left\{e^{-\Hb_f}T_t [\phi(x)\bar\phi(0)]\right\},\\[2mm]
\hat{\D}^{Lf}_{11}(t,\ux)&=&\D^{Lf}(t,\ux)=
(\mbox{Tr}\, e^{-\Hb_f})^{-1}
\mbox{Tr}\left\{e^{-\Hb_f}T_t [e^{iP^0_f t}\phi(0,\ux)e^{-iP^0_f t}
\bar\phi(0)]\right\}\\
&=&D^{Lf}(x')=(\mbox{Tr}\, e^{-\Hb_f})^{-1}
\mbox{Tr}\left\{e^{-\Hb_f}T_+ [\phi(x')\bar\phi(0)]\right\},
\end{eqnarray}
\end{subequations}
where $x'$ is defined in \eq{x'}.
Similar to the zero temperature case, straightforward calculation of 
\eqs{freeprop} shows that
$D^{Lf}(p)=D^f(p)$ [see also \eqs{freespect}], therefore the full propagators 
constructed according to
\eq{ppert} and \eq{LFpert} are equal to each other, $D^{L}(p)=D(p)$,
which already means the equivalence of real time LF and conventional thermal field
theories for the scalar particle case.
As we shall now see, this also leads to the identity between
$\Delta^L(p_0,\up)$ and 
$\Delta(p_0,{\bf p})$, the analytic continuations of the LF and
conventional imaginary time propagators. To define a unique analytic 
continuation of $\Delta(i\omega_n,{\bf p})$, given for
the discrete values
 $\omega_n=2\pi n/\beta$, only two requirements have been needed
in the conventional approach: (i) $|\Delta(z,{\bf p})|\rightarrow 0$ as 
$|z|\rightarrow \infty$, and  (ii) that $\Delta(z,{\bf p})$ is analytic outside the real
axis \cite{lebellac,gorkov,walecka}. Placing the same requirements 
on $\Delta^L(p_0,\up)$,
one obtains a unique analytic continuation of the LF imaginary time propagator
as well. These analytic continuations are provided by \eqs{LehmanLb} and (\ref{Lehmanb}):
\begin{subequations}\label{analyt}
\begin{eqnarray}
\Delta(p_0,{\bf p})&=&-\int\frac{dp'_0}{2\pi}
\frac{\rho(p_0',{\bf p})}{p_0+i\eta-p_0'},\label{analyt1}\\[2mm]
\Delta^L(p_0,\up)&=&-\int\frac{dp'_0}{2\pi}
\frac{\rho^L(p_0',\up)}{p_0+i\eta-p_0'}. \eqn{analyt2}
\end{eqnarray}
\end{subequations}
Equating $D^L(p)$ of \eq{LehmanLa} to $D(p)$ of \eq{Lehmana} leads to the identity of the LF and conventional spectral functions, as well as to the
identity $\Delta^L(p)=\Delta(p)$, as the latter are represented by \eqs{analyt}.

\subsection{Non-scalar particles}

To apply the above proof to the case of non-scalar particles, we note that in the LF
approach the components of the non-scalar field become constrained, and as a result 
of the constaints, the interaction part of the LF Hamiltonian, $P^-_{LI}$,
aquires extra terms, so that  perturbation theory for the LF real time propagator $D^L(x)$ [\eq{LFpert}] has 
extra vertices compared to that for the conventional propagator $D(x)$ [\eq{ppert}]. At the same time, the free non-scalar propagators
in the LF and conventional approaches are different [see \eq{l-f} for the case of a spinor particle]. Without going into details here, the general
strategy to prove equivalence of LF and conventional thermal field theories for the case of non-scalar particles is similar to the one used in the zero temperature case
\cite{equival,rohrlich,chang1,chang2,yan,tomboulis,eyck}. For example, in models of
spin 1/2 fermions where interactions are given by three-point scalar or pseudoscalar vertices, equivalence at zero temperature is proved by showing that the extra vertices in LF perturbation theory can be taken into account by simply adding the term $i\gamma^+/2p^+$ to the free LF propagator, with the sum being equal to the conventional propagator.\footnote{This applies only to internal propagators (those not corresponding to an external leg).} For the same models at non-zero temperature one follows a similar procedure: the extra vertices in the LF perturbation theory of \eq{LFpert} are taken into account by
adding the following instantaneous term to the free LF propagator of \eq{l-f}:
\begin{eqnarray}\label{inst}
i\frac{\gamma^+}{2p^+} 
\left( \begin{array}{cc} 1 & 0  \\
 0 & -1 \end{array} \right).
\end{eqnarray}
This turns the LF propagator into the conventional $2\times 2$ real time propagator. 
In this way one can prove the equivalence of LF and conventional real time thermal field theories for non-scalar particles described by
 Lagrangians for which such an equivalence
has been shown at zero temperature, as done for example in Refs.\
\cite{equival,rohrlich,chang1,chang2,yan,tomboulis,eyck}.\footnote{In this respect it should be noted that vector particles need a more sophisticated treatment \cite{chang2,yan}, and equivalence may not mean that the dressed LF propagator is equal to the conventional one as in the scalar particle case, or even effectively equal to the conventional one as in the spinor case discussed above.} 

 The rest of the proof, showing equivalence of the LF and conventional imaginary time thermal field theories, is 
based on spectral  representations and follows the same procedure presented above for the
scalar particle case.

\section{Free propagators}

With the help of spectral functions, we will derive the free propagators of LF thermal field
theory,  including the spin $1/2$ fermion (spinor) propagator in the entire Dirac index space.
In Ref.\ \cite{das} it was suggested that the fermion propagator be expressed in
the spinor subspace projected by
$P^+=\gamma^-\gamma^+/2$. Yet there is clear need for the fermion propagator 
in the entire spinor space, for example, to keep under control the compensation 
between differences in vertices and propagators with respect 
to the conventional approach.

Using the eigenstates of the noninteracting system in \eq{specfunc}, 
one gets the spectral functions of the free scalar and spinor particles:
\begin{subequations}\label{freespect} 
\begin{eqnarray}
\rho^{Lf}_0(p)&=&\rho^f_0(p)=2\pi\epsilon(p_0)\delta(p^2-m^2)
\hspace{27mm}\mbox{(scalar)},\\[1mm]
\rho^{Lf}_{1/2}(p)&=&\rho^f_{1/2}(p)=2\pi\epsilon(p_0)(p\llap/+m)\delta(p^2-m^2)
\hspace{1cm}\mbox{(spinor)}, \eqn{freespectb}
\end{eqnarray}
\end{subequations}
where $\epsilon(p_0)=p_0/|p_0|$. \eqs{freespect} can also be obtained from \eq{specfun} by using the well known free field commutators/anticommutators.
Although the free spectral functions are the same in the LF and conventional 
approaches, the fermion bare propagators are not:
\begin{subequations}
\begin{eqnarray}\label{scalferm}
D^{Lf}_{1/2}(p)&=&(\bar p\llap/+m)\left[
\frac{i}{p^2-m^2+i\eta}-2\pi f(|p_0|)\delta(p^2-m^2)\right] \eqn{DLF}\\[2mm]
D^f_{1/2}(p)&=&(p\llap/+m)\left[
\frac{i}{p^2-m^2+i\eta}-2\pi f(|p_0|)\delta(p^2-m^2)\right]
\end{eqnarray}
\end{subequations}
where $\bar p$ in \eq{DLF} is the on 
mass shell momentum with components $\bar p^-=(p_\perp^2+m^2)/2p^+$, and
$\underline{\bar p}=
\up$, which depend only on $p^+$ and $p^\perp$.
This difference arises because different components of the four-momentum
$p$ are fixed in the integrals of
\eqs{LehmanL} and (\ref{Lehman}). A similar difference arises  in
the imaginary time formalism where the spinor propagator can be derived
using the same spectral function, given by \eq{freespectb}, and the 
representation
of \eq{LehmanLb}:
\be
\Delta^{Lf}_{1/2}(i\omega_n,\up)=- \,\frac{\bar p\llap/+m}{2p^-_np^+-m^2-p_\perp^2}
\eqn{projprop}
\ee
where $p^-_n=\sqrt{2}i (2n+1)\pi T-p^+=\sqrt{2} p^0_n-p^+$.
When fermions are involved the interaction
part of the LF Hamiltonian has an extra term resulting from resolving 
constraints, whose effect is an instantaneous 
addition of $\gamma^+/2p^+$ to the propagator $\Delta^{Lf}_{1/2}$:
\be
\frac{\bar p\llap/+m}{2p^-_np^+-m^2-p_\perp^2}+\frac{\gamma^+}{2p^+}=
\frac{2p^+(\gamma^+p^-_n+\gamma^-p^+-\gamma_\perp p_\perp+m)}
{2p^+(2p^-_np^+-m^2-p_\perp^2)}=
\frac{p\llap/_n+m}{p^2_n-m^2}\eqn{compens}
\ee
where $\up_n=\up$.
\eq{compens} should be compared with the conventional imaginary time
fermion propagator
\be
\Delta^f_{1/2}(i\omega_n,{\bf p})=\left.\frac{p\llap/+m}{p^2-m^2}\right|_{p_0=i(2n+1)\pi T}
^{ ({\bf p}\ \mbox{\ssz fixed})}=
-\frac{i(2n+1)\pi T\gamma_0-{\bf p}\cdot\bfg+m}
{[(2n+1)\pi T]^2+{\bf p}^2+m^2}. \eqn{convferm1}
\ee
The difference between \eq{compens} and \eq{convferm1} is similar
to the scalar particle case in that the replacement 
$p^0\rightarrow i\pi (2n+1)T$ in the zero temperature propagator is carried 
out when different remaining variables $\up$ and
${\bf p}$ are fixed respectively.

It is not dificult to derive a spectral representation for the $2\times 2$
propagators of the real time formalism and to write down the analog of
\eq{specfunc} for the corresponding $2\times 2$ spectral function. 
Then again we will see that the spectral function of a free particle is
identical for the LF and conventional approach. As a result we will obtain
the following expression for the LF real time fermion propagator
\begin{eqnarray}\label{l-f}
\hat D^{Lf}_{1/2}(p)&=&(\bar{p}\llap/+m) 
\left( \begin{array}{cc} \ds \frac{i}{p^2-m^2+i\eta} & 0  \\
 0 & \ds\frac{-i}{p^2-m^2-i\eta} \end{array} \right)\nn[2mm]
&-&2\pi (p\llap/+m) 
\left( \begin{array}{cc} f(|p_0|) & f(|p_0|)-\theta(-p_0)   \\
 f(|p_0|)-\theta(p_0) & f(|p_0|) \end{array} \right)\delta(p^2-m^2)
\end{eqnarray}
which differs from the conventional one in having $\bar p$ as the on mass shell momentum in the numerator of the first term on the right hand side of the equation.
\newpage

\appendix

\section{Light front quantization}

As the precise meaning of "LF quantization" appears to vary in the current literature, here we give the exact sense in which this term is used in the present paper. It is sufficient to consider only the scalar particle case.

We follow the traditional formulation of LF quantization.
Starting with a Lagrangian density $\mathscr{L}[\Phi]\equiv \mathscr{L}(\Phi(x),\partial_\mu\Phi(x))$ and the equations of motion
\be
\frac{\partial}{\partial x^\mu} 
\frac{\partial \mathscr{L}}{\partial(\partial\Phi/\partial x^\mu)}    \eqn{emotion}
-\frac{\partial \mathscr{L}}{\partial\Phi}=0,
\ee
quantization enters through the specification of the initial conditions for these equations. In the case of LF quantization, we specify the initial conditions as
\be
\Pi_L(x)=\frac{\partial \mathscr{L}}{\partial(\partial\Phi_L/\partial x^+)},\hspace{1cm}
[\Phi_L(0),\Pi_L(x)]\delta(x^+)=\frac{i}{2}\delta^4(x)  \eqn{LFquant}
\ee
where $\Pi_L(x)$ is the momentum conjugate to the field $\Phi_L(x)$, and the commutation relation between $\Pi_L$ and $\Phi_L$ is specified on the  LF surface $x^+=0$. Note that we have added a subscript $L$ to those solutions of \eq{emotion} whose initial conditions are given by \eq{LFquant}. The LF initial conditions should be contrasted with those of usual equal time quantization:
\be
\Pi(x)=\frac{\partial \mathscr{L}}{\partial(\partial\Phi/\partial x^0)},\hspace{1cm}
[\Phi(0),\Pi(x)]\delta(x^0)= i \delta^4(x) . \eqn{quant}
\ee

The LF four-momentum $P_L^\mu$ is given as an integral on the same surface as that specifying  the LF commutation relations:
\be
P_L^\mu=\int \delta(x^+)\, T^{+\mu}[\Phi_L] \,d^4x=\int_{x^+=0}  T^{+\mu} [\Phi_L]\,d{\ux}
\eqn{PL_def}
\ee
where the energy-momentum tensor $T^{\mu\nu}$ is connected to the Lagrangian in a
way that is independent of quantization:
\be
T^{\mu\nu}[\Phi]=\frac{\partial \mathscr{L}}{\partial(\partial\Phi/\partial x^\mu)} \frac{\partial \Phi}{\partial x_\nu}-g^{\mu\nu} \mathscr{L}[\Phi].
\ee
Similarly, the conventional four-momentum is defined by
\be
P^\mu=\int \delta(x^0)\, T^{0\mu}[\Phi] \,d^4x=\int_{x^0=0}  T^{0\mu} [\Phi]\,d{\bf x}.
\eqn{P_def}
\ee
In the free case (no interactions), one can show that the LF and conventional free fields are identical, as are the LF and conventional free momenta:
\be
\phi_L(x) = \phi(x),\hspace{1.5cm} P^\mu_{Lf} = P^\mu_f.
\ee
Constructed in this way, both the LF and conventional four-momenta act as generators of
space-time translations:
\begin{subequations}
\begin{align}
\left[P_L^\mu,\Phi_L(x)\right]=-i\frac{\partial\Phi_L(x)}{\partial x_\mu}, \\[2mm]
\left[P^\mu,\Phi(x)\right]=-i\frac{\partial\Phi(x)}{\partial x_\mu}.
\end{align}
\end{subequations}
We note the universal nature of these commutation relations -  they do not depend on the type of quantization used.
For the LF energy operator, $P_L^0$, and the LF Hamiltonian, 
$P_L^-=\frac{1}{\sqrt{2}}(P_L^0-P_L^3)$, we have that
\begin{align}
\left[P^0_L,\Phi_L(x)\right]& =\left. -i\frac{\partial\Phi_L(x)}{\partial x^0}\right|_{x^3,x^\perp\,\mbox{\ssz fixed}}=
\left. \frac{-i}{\sqrt{2}}\frac{\partial\Phi_L(x)}
{\partial x^+}\right|_{x^3,x^\perp\,\mbox{\ssz fixed}} , \eqn{trans1} \\[2mm]
\left[P^-_L,\Phi_L(x)\right]&=\left. -i\frac{\partial\Phi_L(x)}
{\partial x^+}\right|_{x^-,x^\perp\,\mbox{\ssz fixed}}=\left.
\frac{-i}{\sqrt{2}}\frac{\partial\Phi_L(x)}
{\partial x^0}\right|_{x^-,x^\perp\,\mbox{\ssz fixed}}  .  \eqn{trans2} 
\end{align}
It is seen that both $P^0_L$ and $P^-_L$ determine evolution in LF time $x^+$, the only difference being in the variables which are kept constant.

As the LF four-momentum $P^\mu_L$ is defined on  the LF hyperplane $x^+=0$, as indicated by
\eq{PL_def}, it follows that it is a functional of $\Phi_L(0,\ux)=\phi(0,\ux)$. With the same being true
of the free momentum operator $P^\mu_f$, it follows that 
$P_{LI}^0(\tau)\equiv e^{\tau P^0_f}(P^0_L-P^0_f)e^{-\tau P^0_f}$  depends
on $\phi_{\tau}(\ux)=e^{P^0_f\tau}\phi(0,\ux)e^{-P^0_f\tau}$. This observation is at the heart of LF imaginary time perturbation theory, discussed in Appendix B.

\section{LF imaginary time perturbation theory}

Here we give a brief derivation of perturbation theory for the fully dressed LF
imaginary time propagator $\Delta^L$ given in \eqs{danger}. The perturbation theory is based on the following operator exponent expansion
\be
\mathscr{U}(\tau_1, \tau_2)\equiv e^{ P^0_f \tau_1}e^{P^0_L(\tau_2-\tau_1)}e^{-P^0_f\tau_2 }=
T_\tau \exp\left[ \int_{\tau_1}^{\tau_2} P_{LI}^0(\tau)d\tau\right]
\ee
where $P^0_L$ and $P^0_f$ are the zero components of the LF four-momentum operator in the
interacting and free case, respectively,
$P_{LI}^0(\tau)=e^{P^0_f\tau}(\HL -P^0_f)e^{-P^0_f\tau}$, and $T_\tau$ is the ordering operator, as in \eq{tauord}, which
orders $\tau$-dependent quantities with respect to $\tau$, so that
\be
T_\tau [P^0_{LI}(\tau)P^0_{LI}(\tau')]=\theta(\tau-\tau')P^0_{LI}(\tau)P^0_{LI}(\tau')+
\theta(\tau'-\tau)P^0_{LI}(\tau')P^0_{LI}(\tau) . \eqn{tauorder}
\ee
Note that  both $P^0_L$ and $P^0_f$ are expressed in terms of the free
 field operators $\phi(0,\ux)$ on the LF hyperplane $x^+=0$.
Then by analogy with textbook derivations of perturbation theory in the imaginary time formalism
\cite{lebellac,gorkov,walecka}, the Green function $\Delta^L$ of \eq{dangera} can be written in the form 
\begin{eqnarray}\label{LFpertheo}
&&(\mbox{Tr}\,e^{-\beta \HL })\Delta^L(\tau,\ux)=
\mbox{Tr}\left\{e^{-\beta \HL }T_\tau [e^{\HL \tau}\phi(0,\ux)
e^{-\HL \tau}\bar\phi(0)]\right\}\nn
&=&
\mbox{Tr}\left\{e^{-\Hb_f}T_\tau \left[e^{\int_{0}^{\beta} 
P_{LI}^0(\tau')d\tau'}e^{P^0_f\tau}\phi(0,\ux)
e^{-P^0_f\tau}\bar\phi(0)\right]\right\}
\end{eqnarray}
where $T_\tau$ orders operators $P^0_{LI}(\tau')$, 
$\phi_\tau(\ux)=e^{P^0_f\tau}\phi(0,\ux)e^{-P^0_f\tau}$ and 
$\bar\phi_0(\underline{0})=\bar\phi(0)$ with respect to $\tau'$, $\tau$ 
and $0$ according \eq{tauord}. It is important to note that as a result of
the definition of  \eq{dangera}, the interaction part of the energy operator,
$P_{LI}^0(\tau')$, depends on
$\phi_{\tau'}(\ux)=e^{P^0_f\tau'}\phi(0,\ux)e^{-P^0_f\tau'}$, the free 
field operators on the $x^+=0$ LF hyperplane $\phi_0(\ux)=\phi(0,\ux)$, shifted by imaginary usual time. 
In the analogous
conventional expression given in Refs.\  \cite{lebellac,gorkov,walecka},
the interaction part of the energy depends on the free field operators on 
the hyperplane $t=0$ (not $x^+=0$) shifted by imaginary usual time. This 
difference is reflected in the difference of the imaginary time  propagators of the 
corresponding perturbation theories. Indeed using Wick's theorem in \eq{LFpertheo}
one ends up with imaginary time perturbation theory with propagators
\begin{eqnarray}\label{frimpeop}
\Delta^{Lf}(\tau,\ux)&=&(\mbox{Tr}\,e^{-\Hb_f})^{-1}
\mbox{Tr}\left\{e^{-\Hb_f}T_\tau [e^{P^0_f\tau}\phi(0,\ux)
e^{-P^0_f\tau}\bar\phi(0)]\right\}\\
\Delta^{Lf}(i\omega_n,\up)&=& \frac{1}{\sqrt{2}}
\int_0^\beta d\tau d\ux\, e^{i(\omega_n\tau-\up\cdot\ux)}\,
\Delta^{Lf}(\tau,\ux)
\end{eqnarray}
Applying the arguments leading to \eq{x'} to the imaginary time $t=-i\tau$, one gets
$e^{P^0_f\tau}\phi(0,\ux)e^{-P^0_f\tau}=\phi(z)$, where $z$ is a complex coordinate space
four-vector with purely imaginary LF time $z^+=-i\tau/\sqrt{2}$ and real
spatial components $z^3=-x^-/\sqrt{2}$ and $z^\perp=x^\perp$.
This suggests that the propagator of \eq{frimpeop}
corresponds to imaginary LF time ordering, in contrast to the usual time 
ordering in the conventional approach.
Direct calculation leads to the scalar particle imaginary time propagator 
suggested in Ref.\  \cite{das},
\be
\Delta^{Lf}_0(i\omega_n,\up)= - \frac{1}{2p^-_np^+-m^2-p_\perp^2},
\eqn{scimprop}
\ee
with $p^-_n=\sqrt{2}i\pi 2nT-p^+=\sqrt{2} p^0_n-p^+$,
and the spinor particle propagator of \eq{projprop} derived above. Even in the scalar 
particles case, the difference between the imaginary time formalisms in the two quantisations is more
than a simple change of variables: although the energy variable is purely 
imaginary in both \eq{scimprop} and the conventional propagator, $p^3$ is real
in the conventional approach but complex in the LF one.  
Note also that in the LF dynamics the interaction does not affect
$P^+_L$. Only the operator $P^-_L$  acquires an interaction part, therefore
the interaction part of the energy operator is $P^0_{LI}=P^-_{LI}/\sqrt{2}$
and in the case of scalar particles (only) it is related to the interaction
Lagrangian (without coupling with derivatives) by a factor of $\sqrt{2}$:
$P^0_{LI}=-L_I/\sqrt{2}$
\section{Light-front spectral function}

Here we introduce the LF spectral function $\rho^L$ used in \eqs{LehmanL},  and specify some of its properties by exploiting the analogy with the well 
known spectral function $\rho$ of conventional thermal field theory \cite{lebellac}. In coordinate 
space, one can use \eq{DLa}, defining the real time LF propagator $D^L(x)$, to write
\be
D^L(x)=\theta(x^+)D^{L>}(x)\pm \theta(-x^+)D^{L<}(x)=\theta(x^+)\rho^L(x)
\pm D^{L<}(x) \eqn{c1}
\ee
where 
\begin{align}
D^{L>}(x)&=(\mbox{Tr}\,e^{-\beta P^0_L})^{-1}
\mbox{Tr}\left\{e^{-\beta P^0_L}\Phi_L(x)\bar\Phi_L(0)\right\},\\[1mm]
D^{L<}(x)&=(\mbox{Tr}\,e^{-\beta P^0_L})^{-1}
\mbox{Tr}\left\{e^{-\beta P^0_L}\bar\Phi_L(0)\Phi_L(x)\right\},
\end{align}
and
\be
\rho^L(x)=D^{L>}(x)\mp D^{L<}(x)=(\mbox{Tr}\,e^{-\beta P^0_L})^{-1}
\mbox{Tr}\left\{e^{-\beta P^0_L}[\Phi_L(x),\bar\Phi_L(0)]_\mp\right\},
\eqn{specfun}
\ee
is the LF spectral function defined as the ensemble average of the commutator / anticommutator
$[\Phi_L(x),\bar\Phi_L(0)]_\mp$.
The Kubo-Martin-Schwinger relation is the same as in Ref.\ \cite{lebellac},
\be
D^{L<}(x)=D^{L>}(x^0-i\beta,{\bf x}).
\ee
It is easy to express the above equations in momentum space thereby obtaining the following relations:
\begin{align}
\rho^L(p)&=D^{L>}(p)\mp D^{L<}(p),\\[1mm]
D^{L<}(p)&=\pm e^{-\beta p^0}D^{L>}(p),\\[1mm]
D^{L>}(p)&=[1\pm f(p^0)]\rho^L(p),\\[1mm]
D^{L<}(p)&=\pm f(p^0)\rho^L(p),
\end{align}
which lead to \eq{LehmanLa}. The $x^+$ ordering in \eq{c1} (instead of
conventional $x^0$ ordering) is reflected 
in that $\up$ is fixed (not ${\bf p}$) in $\rho^L(p)$ in the spectral 
energy integral of \eq{LehmanLa}. To derive \eq{LehmanLb}, one can easily verify that
\be
\Delta^L(\tau,\ux)=\mathscr{D}^{L>}(-i\tau,\ux) = D^{L>}(z)
\ee
with $\ds z^+=-i\frac{\tau}{\sqrt{2}}$, $\ds z^3=-\frac{x^-}{\sqrt{2}}$, and $\ds z^\perp=x^\perp$, 
if $\tau$ is in the interval 
$[0,\beta]$.


\begin{thebibliography}{99}
\setlength{\itemsep}{-.5mm}
\bibitem{brodsky} S. J. Brodsky, Nucl. Phys. Proc. Suppl. {\bf 108}, 327 (2002).
\bibitem{das} V. S. Alves, Ashok Das, and Silvana Perez, Phys. Rev.
D {\bf 66}, 125008 (2002).
\bibitem{weldon1} H. Arthur Weldon, Phys. Rev. D {\bf 67}, 085027 (2003).
\bibitem{weldon2} H. Arthur Weldon, Phys. Rev. D {\bf 67}, 128701 (2003).
\bibitem{equival} J. B. Kogut and D. E. Soper, Phys. Rev. D {\bf 1}, 2901
(1970).
\bibitem{rohrlich} F. Rohrlich, Acta Phys. Austriaca {\bf 32}, 87 (1970).
\bibitem{chang1} S. J. Chang, R. G. Root, and Tung-Mow Yan, 
Phys. Rev. D {\bf 7}, 1133 (1973).
\bibitem{chang2} S. J. Chang and Tung-Mow Yan, 
Phys. Rev. D {\bf 7}, 1147 (1973).
\bibitem{yan} Tung-Mow Yan, Phys. Rev. D {\bf 7}, 1760 (1973).
\bibitem{tomboulis} E. Tomboulis, Phys. Rev. D {\bf 8}, 2736 (1973).
\bibitem{eyck} J. H. Ten Eyck and F. Rohrlich, Phys. Rev. D {\bf 9}, 2237
(1974).
\bibitem{mccartor} G. McCartor and D. G. Robertson, Z. Phys. C{\bf 53}, 679
(1992) .
\bibitem{Alex} S. Elster and A. C. Kalloniatis, Phys.\ Lett.\ B {\bf 375}, 285 (1996).
\bibitem{lebellac} M. Le Bellac, {\em Thermal Field Theory} (Cambridge University
Press, Cambridge, England, 1996).
\bibitem{gorkov} A.Abrikosov, L.Gor'kov, and I.Dzyaloshinsky,
{\it Methods of Quantum Field Theory in Statistical Physics} (Dover Publications,
New York, 1977).
\bibitem{walecka} A. L. Fetter and J. D. Walecka, {\it Quantum Theory of 
Many-Particle Systems} (McGraw Hill, New York, 1971).
\bibitem{smilga} A. V. Smilga, {\it Physics of Thermal QCD},
Phys. Rept. {\bf 291}, 1 (1997). 
\end{thebibliography}
\end{document}